\documentstyle[12pt,psfig]{article}
\newcommand{\be}{\begin{equation}}
\newcommand{\ee}{\end{equation}}

\newcommand{\beqa}{\begin{eqnarray}}
\newcommand{\eeqa}{\end{eqnarray}}
\newcommand{\nn}{\nonumber}

\newcommand{\eqref}[1]{(\ref{#1})}

\def\boxit#1{\vbox{\hrule\hbox{\vrule\kern8pt
\vbox{\hbox{\kern8pt}\hbox{\vbox{#1}}\hbox{\kern8pt}}
\kern8pt\vrule}\hrule}}
\def\mathboxit#1{\vbox{\hrule\hbox{\vrule\kern8pt\vbox{\kern8pt
\hbox{$\displaystyle #1$}\kern8pt}\kern8pt\vrule}\hrule}}

\def\IB{\relax\hbox{$\inbar\kern-.3em{\rm B}$}}
\def\IC{\relax\hbox{$\inbar\kern-.3em{\rm C}$}}
\def\ID{\relax\hbox{$\inbar\kern-.3em{\rm D}$}}
\def\IE{\relax\hbox{$\inbar\kern-.3em{\rm E}$}}
\def\IF{\relax\hbox{$\inbar\kern-.3em{\rm F}$}}
\def\IG{\relax\hbox{$\inbar\kern-.3em{\rm G}$}}
\def\IGa{\relax\hbox{${\rm I}\kern-.18em\Gamma$}}
\def\IH{\relax{\rm I\kern-.18em H}}
\def\IK{\relax{\rm I\kern-.18em K}}
\def\IL{\relax{\rm I\kern-.18em L}}
\def\IP{\relax{\rm I\kern-.18em P}}
\def\IR{\relax{\rm I\kern-.18em R}}
\def\IZ{\relax\ifmmode\mathchoice
{\hbox{\cmss Z\kern-.4em Z}}{\hbox{\cmss Z\kern-.4em Z}}
{\lower.9pt\hbox{\cmsss Z\kern-.4em Z}} {\lower1.2pt\hbox{\cmsss
Z\kern-.4em Z}}\else{\cmss Z\kern-.4em Z}\fi}

\def\II{\relax{\rm I\kern-.18em I}}

\def\CM {{\cal M}}

\def\CV {{\cal V}}

\begin{document}

\begin{flushright}
{\tt NRCPS-HE-56-10} \\
\end{flushright}

\vspace{1cm}
%\begin{titlepage}
%\title{
\begin{center}
{\Large ~\\{\it Production of
non-Abelian Tensor Gauge Bosons\\
\vspace{1cm}

Tree Amplitudes and BCFW Recursion Relation

}

}%title ends

\vspace{1cm}
%\author{

{ \it{George Georgiou} }
and
{ \it{George  Savvidy  } }\\
\vspace{1cm}
{\sl Demokritos National Research Center\\
Institute of Nuclear Physics\\
Ag. Paraskevi, GR-15310 Athens,Greece  \\
\centerline{\footnotesize\it E-mail: georgiou@inp.demokritos.gr, savvidy@inp.demokritos.gr}
}%author ends
%}
%\date{}%in order NOT to write the date
%\maketitle
\end{center}
\vspace{60pt}

\centerline{{\bf Abstract}}
The BCFW recursion relation is used to calculate tree-level scattering amplitudes
in generalized Yang-Mills theory and, in particular, four-particle amplitudes
for the production rate of non-Abelian tensor gauge bosons of arbitrary
high spin  in the fusion of two gluons.
The consistency of the calculations in different kinematical channels is fulfilled when all
dimensionless cubic coupling constants between vector bosons and high spin non-Abelian tensor gauge bosons
are equal to the Yang-Mills coupling constant.
We derive a generalization of the Parke-Taylor formula in the case
of production of two tensor gauge bosons of spin-s and N gluons (jets).
The expression is holomorhic in the spinor variables of the scattered particles, exactly as
the MHV gluon amplitude is, and reduces to the gluonic MHV amplitude when s=1.

\vspace{12pt}

\noindent

\newpage

%\end{abstract}
%\thispagestyle{empty}
%\end{titlepage}

%\tableofcontents

\pagestyle{plain}
%\pagenumbering{roman}

\section{\it Introduction}

The Lagrangian of non-Abelian tensor gauge fields describes the interaction of
the Yang-Mills quanta with massless  tensor gauge bosons of increasing
spins \cite{Savvidy:2005fi,Savvidy:2005zm,Savvidy:2005ki,Savvidy:2010vb}. The characteristic property of
generalized Yang-Mills theory is that all interaction vertices
between Yang-Mills and high-spin fields have {\it dimensionless coupling constants} in four-dimensional space-time.
That is, the cubic interaction vertices have only first order derivatives and the quartic vertices
have no derivatives at all.

One of the first calculations of tree level scattering amplitudes
in generalized Yang-Mills theory was made in a series of articles
\cite{Konitopoulos:2008je,Konitopoulos:2008bd}
where the authors considered the creation of tensor gauge bosons in the annihilation
processes of quarks and gluons.
The main problem in this calculation was the necessity to sum over infinitely many diagrams even in the
lowest order of the perturbation theory because the kinetic term
of the Lagrangian contains non-diagonal transitions.
Our intension in this article is to use a different technique to resolve these difficulties.

Recently, a very powerful technique  was developed  for the calculation of high order tree level diagrams in
Yang-Mills and other supersymmetric theories \cite{Berends:1981rb,Kleiss:1985yh,Xu:1986xb,Gunion:1985vca,Dixon:1996wi,Parke:1986gb,Berends:1987me,
Witten:2003nn,Britto:2004ap,Britto:2005fq,Benincasa:2007xk,Cachazo:2004kj,Georgiou:2004by,Georgiou:2004wu}.
It uses spinor representation of the
scattering amplitudes and dramatically simplifies the calculations
\cite{Dixon:1996wi,Britto:2004ap,Britto:2005fq,Benincasa:2007xk}. The  advantage of
this approach is that it allows the computation of high order scattering amplitudes in terms of lower ones,
expressing any tree amplitude as a sum over terms constructed
from products of two amplitudes of fewer particles multiplied by a Feynman propagator.
The two amplitudes in each term are physical, in the sense that all particles are on-shell and
momentum conservation is preserved \cite{Britto:2004ap,Britto:2005fq,Benincasa:2007xk}.
Here the use of the complex momenta allows
to write non-vanishing three-particle on-shell vertices as well as to deform two
of the momenta in an arbitrary scattering amplitude along a complex direction
defined by the deformation parameter $z$.
Any tree-level amplitude becomes a rational function of the complex parameter z,
with at most simple poles and if the amplitude vanishes at large z, then
it can be computed by knowing the position of the poles and the value of the residues.
These are on-shell data that are completely specified by the three-point on-shell vertices.

The application of the BCFW recursion relation to calculate four-particle amplitudes
allows to derive the production rate of non-Abelian tensor gauge bosons of arbitrary
high spin in the fusion of two gluons $G+G \rightarrow T+T$.
The consistency of the calculations in different kinematical channels is fulfilled when all
cubic coupling constants between vector bosons (gluons) and high spin tensor bosons
are of the generalized Yang-Mills type \cite{Savvidy:2005fi,Savvidy:2005zm,Savvidy:2005ki,Savvidy:2010vb}
and are equal to the Yang-Mills coupling constant
\be
g_{1ss} = g_{1ss-2}= g_{YM},~~~~~~s= 2,3,....
\ee
We have checked that the amplitude vanishes quickly enough as deformation parameter $z$ tends to infinity,
so that there is no contribution from the contour at infinity.
The result can be expressed in a compact form
\beqa\label{1}
d\sigma_{ +s}  =
 ~\biggl({1-\cos\theta \over  1+ \cos\theta}\biggr)^{2s-2}
d\sigma_{ +1 },  ~~~s=1,2,3,...
\eeqa
where $d\sigma_{ +1 }$ is polarized cross section of two gluons into two gluons $G+G \rightarrow G+G$
and $\theta$ is the scattering angle.
The formula demonstrates the complete dependence of the cross section on the spin of the  tensor
gauge bosons and  allows to sum contributions from all spins.

We derive the four-particle cross section for the scattering of high spin tensor gauge bosons
of the helicities $h_1,h_2,h_3,h_4$ ($h_1+h_2+h_3+h_4=0,~h_1>0,h_2<0$), which has the following form:
\beqa\label{crosssectiongeneral0}
d\sigma_{h_1,h_2,h_3,h_4}  =
 {(1-\cos\theta)^{2h_1+2h_2+2h_3-2} \over 2^{2h_1-2h_2 -4} (1+ \cos\theta)^{2h_3-2} }
d\sigma_{ +1 }
 \eeqa
and {\it spectacularly falls exponentially as the spins of scattering particles increases}.

We also derive a generalization of the Parke-Taylor formula in the case
of production of two tensor gauge bosons of spin-$s$ and $(n-4)$ gluons (jets)
in the amplitude $G+G \rightarrow T+T +(n-4)~ G$. The result reads:
\be\label{park0}
M^{(n)}(1^+,...i^-, ...,k^{+s},....j^{-s},...,n^+)=
g^{n-2} \frac{<ij>^4}{\prod_{l=1}^{n} <l l+1>} \Big( \frac{<ij>}{<ik>}\Big)^{2s-2},
\ee
where $n$ is the total number of particles, and the dots stand for the
positive helicity gluons. Furthermore, $i$ is the position of the negative helicity gluon,
while  $k$ and $j$ are the positions of the particles with helicities $+s$ and $-s$ respectively.
This expression is holomorhic in the spinors of the particles, exactly as
the MHV gluon amplitude and for $s=1$ the second fraction in \eqref{park0}
is absent and \eqref{park0} reduces to the well-known result for the MHV amplitude \cite{Parke:1986gb}.

In the next section, we shall review the spinor representation of scattering amplitudes. In the third section,
we discuss the three-point on-shell vertices for complex momenta
and in the forth section we describe
a class of the three-point on-shell vertices which have dimensionless coupling constants corresponding
to cubic vertices of the generalized Yang-Mills theory. In the fifth section, we apply the Benincasa-Cachazo
recursion relation to compute four-particle scattering amplitude of gluons and
tensor gauge bosons using three-point on-shell vertices of the generalized Yang-Mills theory.
The production cross sections (\ref{1}) of the non-Abelian
tensor gauge bosons of arbitrary spin-s and the general formula (\ref{crosssectiongeneral0})
are derived in the sixth section and the generalization of the
Parke-Taylor formula (\ref{park0}) in the last, seventh section.

\section{\it Scattering Amplitudes in Spinor Representation}

Let us consider a scattering amplitude for
massless particles of momenta $p_i$ and polarization tensors $\varepsilon_i$ ~(i=1,...,n),
which are described by irreducible massless representations of Poincar\'e group and are
classified by their helicities $h= \pm s$, where $s$ is an integer,
\be\label{smatrix}
M_n = M_n(p_1,\varepsilon_1;~p_2,\varepsilon_2;~...;~p_n,\varepsilon_n).
\ee
We are interested in representing the momenta $p_i$ and polarization tensors $\varepsilon_i$ in terms of spinors
and the above scattering amplitude in terms of rational functions of spinor products
\cite{Berends:1981rb,Kleiss:1985yh,Xu:1986xb,Gunion:1985vca,Dixon:1996wi,Parke:1986gb,Berends:1987me,
Witten:2003nn,Britto:2004ap,Britto:2005fq,Benincasa:2007xk,Cachazo:2004kj,Georgiou:2004by,Georgiou:2004wu,ArkaniHamed:2008yf}.

The spinor representation of momenta $p^{\mu}$ and
polarization tensors $\varepsilon_i$ can be constructed as follows.
The spinors $\{ \lambda_a, \tilde{\lambda}_{\dot{a}}  \}$ transform in the
representation $(1/2,0)$ and $(0,1/2)$ of the universal cover of the Lorentz
group, $SL(2,C)$, respectively. Invariant tensors are $\epsilon^{ab}$,
$\epsilon^{\dot{a}\dot{b}}$ and $(\sigma^{\mu})_{a\dot{a}}$, where
$\sigma^{\mu}= (1,\vec{\sigma})$. The basic Lorentz invariant spinor products can be
constructed as follows:
$$
\lambda_a \lambda'_b \epsilon^{ab}  \equiv <\lambda,\lambda'>,~~~~~
\tilde{\lambda}_{\dot{a}} \tilde{\lambda}'_{\dot{b}} \epsilon^{\dot{a}\dot{b}}  \equiv [\lambda,\lambda'].
$$
The scalar product of two vectors $p^{\mu}$ and $q^{\nu}$ is given by the product
$$
2 (p\cdot q) = <\lambda^p, \lambda^q> [\tilde{\lambda}^p, \tilde{\lambda}^q].
$$
Using the third invariant tensor one can define
$p^{\mu} = \lambda^a (\sigma^{\mu})_{a\dot{a}} \tilde{\lambda}^{\dot{a}}$ and find out the
corresponding spinor representation of massless particle momentum in the form
\be\label{momentum}
p_{a\dot{a}}= \lambda_a \tilde{\lambda}_{\dot{a}}.
\ee
The corresponding polarization vectors of spin-1 particles are given by
\be\label{poltensors}
\varepsilon^{-}_{a\dot{a}}(p) = {\lambda_a \tilde{\mu}_{\dot{a}} \over [\tilde{\lambda},\tilde{\mu} ]},~~~
\varepsilon^{+}_{a\dot{a}}(p) = {\mu_a \tilde{\lambda}_{\dot{a}} \over <\mu,\lambda >}
\ee
with $\mu_a$ and $\tilde{\mu}_{\dot{a}}$ as arbitrary reference spinors.
One can check that in this representation the following basic properties of polarization tensors are
fulfilled:
\beqa
&p \cdot \varepsilon^{-}(p) = p \cdot \varepsilon^{+}(p)=0 \nn \\
&\varepsilon^{-}(p)\cdot \varepsilon^{-}(p) = \varepsilon^{+}(p)\cdot \varepsilon^{+}(p)=0 \nn\\
&\varepsilon^{+}(p) \cdot \varepsilon^{-}(p) = - 1\nn
\eeqa
The polarization tensors of massless particles of integer spin s can be expressed in terms
of spin-1 particle as follows\footnote{ In labeling helicities, we consider all particles to be outgoing.}:
\beqa
\varepsilon^{-}_{a_1\dot{a}_1,...,a_s\dot{a}_s }(p) = \prod^{s}_{i=1} \varepsilon^{-}_{a_i\dot{a}_i},~~~~~~~~
\varepsilon^{+}_{a_1\dot{a}_1,...,a_s\dot{a}_s }(p) = \prod^{s}_{i=1} \varepsilon^{+}_{a_i\dot{a}_i}.
\eeqa
The presence in (\ref{poltensors}) of  arbitrary reference spinors $\mu_a$ and $\tilde{\mu}_{\dot{a}}$
means that polarization tensors are not uniquely fixed once $\{\lambda, \tilde{\lambda},h\} $ is given.
Let us  consider the transformation
$$
\tilde{\mu}_{\dot{a}} \rightarrow \tilde{\mu}_{\dot{a}} + \eta \tilde{\mu}_{\dot{a}} +
\eta' \tilde{\lambda}_{\dot{a}}.
$$
The action of this transformation on the polarization vector $\varepsilon^{-}_{a\dot{a}}(p)  $ can
be computed:
$$
\varepsilon^{-}_{a\dot{a}}  \rightarrow {\lambda_a (\tilde{\mu}_{\dot{a}} + \eta \tilde{\mu}_{\dot{a}} +
\eta' \tilde{\lambda}_{\dot{a}}) \over [\tilde{\lambda},\tilde{\mu}  + \eta \tilde{\mu}  +
\eta' \tilde{\lambda}  ]} = \varepsilon^{-}_{a\dot{a}}  +
{ \eta'  \over (1+ \eta) [\tilde{\lambda}, \tilde{\mu} ]}\lambda_a \tilde{\lambda}_{\dot{a}}=
\varepsilon^{-}_{a\dot{a}}  + w ~p_{a\dot{a}},
$$
where $$w = { \eta'  \over (1+ \eta) [\tilde{\lambda}, \tilde{\mu} ]},~~~~~
\lambda_a \tilde{\lambda}_{\dot{a}}=p_{a\dot{a}}$$
and therefore represents a  gauge transformation.

Thus the scattering amplitude  of massless bosons $M_n$ (\ref{smatrix}) can  be considered
as a function of spinors $\lambda_i$, $\tilde{\lambda}_i$ and helicities $h_i$:
\be
M_n=M_n(\lambda_1,\tilde{\lambda}_1,h_1;~...;~\lambda_n,\tilde{\lambda}_n,h_n).
\ee
The Poincar\'e invariance requires that $M_n$ gives the same answer independently
of the choice of reference spinors $\mu_i$ \cite{Weinberg:1964cn,Weinberg:1964ev,Weinberg:1964ew}.

Let us consider scaling transformation of spinors which leaves the momenta in (\ref{momentum}) intact:
$$
\lambda \rightarrow u \lambda,~~~~~\tilde{\lambda} \rightarrow u^{-1} \tilde{\lambda},
$$
then
$$
\varepsilon^{-}_{a\dot{a}}  \rightarrow u^2 \varepsilon^{-}_{a\dot{a}} ,~~~~~~~
\varepsilon^{+}_{a\dot{a}}  \rightarrow u^{-2} \varepsilon^{+}_{a\dot{a}}
$$
and the scattering amplitude $M_n$ should be a homogeneous function of these spinors of order $2 h_i$
\cite{Witten:2003nn}:
\be\label{helicityequation}
(\lambda^{a}_{i} {\partial \over \lambda^{a}_{i} }  -
\tilde{\lambda}^{a}_{i} {\partial \over \tilde{\lambda}^{a}_{i} })
M_n(...~\{\lambda_i,\tilde{\lambda}_i,h_i\}~...)= -2 h_i ~M_n(...~\{\lambda_i,\tilde{\lambda}_i,h_i\}...).
\ee
From this equation one can derive a general structure of the three-particle amplitudes $M_3$ in spinor
representation \cite{Benincasa:2007xk}. This will provide us with unique information
about the structure of three-point on-shell vertices $\CV_3(p_1,p_2,p_3)$ in high spin quantum field theory.

\section{\it Three-Point On-Shell Vertices }
In the generalized Yang-Mills theory
\cite{Savvidy:2005fi,Savvidy:2005zm,Savvidy:2005ki,Savvidy:2010vb},
the three-point on-shell vertices can be computed
as a product of a three-particle vertex
$\CV_3$ times three polarization tensors $\varepsilon$ \cite{Savvidy:2008ks,Antoniadis:2009rd}:
\be\label{matrixelement3}
M_3(p_1,\varepsilon_1;p_2,\varepsilon_2;p_3,\varepsilon_3)=
\CV_3(p_1,p_2,p_3) ~\varepsilon_1(p_1) ~\varepsilon_2(p_2)~ \varepsilon_3(p_3),
\ee
where the mass-shell conditions and momentum conservation are $p^2_1 =p^2_2=p^2_3 =0$ and $p_1 + p_2 +p_3 =0$.
The three point scattering amplitude for on-shell massless particles
is equal to zero for the real momenta $p_i$, but if one allows complex momenta or a
different space-time signature
\cite{Witten:2003nn,Britto:2004ap,Britto:2005fq,Benincasa:2007xk,Savvidy:2008ks,Antoniadis:2009rd}
then these matrix elements will have nontrivial behavior.
Indeed, the three-point on-shell vertices are naturally non-zero if one chooses to work
with complex momenta $p^{\mu}$  \cite{Witten:2003nn,Benincasa:2007xk}. In that case   $(\lambda_a)^*  \neq
\pm \tilde{\lambda}_{\dot{a}}$, and the spinors of each particle, $\lambda_i$ and
$\tilde{\lambda}_i$, are independent vectors in $C^2$.

Momentum conservation  and  mass-shell conditions
imply that $p_i \cdot p_j =0$, that is, in spinor representation we have
\be\label{collinearcondition}
<1,2>[1,2] = 0,~~~<2,3>[2,3] = 0,~~~<3,1>[3,1] = 0
\ee
and either spinors $\lambda_i~ (i=1,2,3)$  or spinors $\tilde{\lambda}_i~ (i=1,2,3)$ are collinear.
This means that the non-trivial solution of (\ref{collinearcondition}) is
either $$<1,2>=<2,3> = <3,1> = 0$$ or $$[1,2]=[2,3] = [3,1] = 0.$$
Therefore the three-point on-shell vertex, $M_3(...~\lambda_i,\tilde{\lambda}_i,h_i~...)$,
which is restricted to be a function of $<i,j>$ and $[i,j]$, splits into two independent parts
\cite{Benincasa:2007xk}
$$
M_3 = A(<1,2>,<2,3>,<3,1>)~ + ~B(~[1,2],~[2,3],~[3,1]~).
$$
As a result, equation (\ref{helicityequation}) also splits into two equations for
amplitudes $A$ and $B$
\beqa
 \lambda^{a}_{i} {\partial \over \lambda^{a}_{i} }~
A(<1,2>,<2,3>,<3,1>)= -2 h_i ~A(<1,2>,<2,3>,<3,1>)\nn
\eeqa
and
\beqa
\tilde{\lambda}^{a}_{i} {\partial \over \tilde{\lambda}^{a}_{i} }
B(~[1,2],~[2,3],~[3,1]~)=  2 h_i ~B(~[1,2],~[2,3],~[3,1]~).\nn
\eeqa
Looking for a polynomial solution of the first equation in the form
$<1,2>^{d_3} <2,3>^{d_1} <3,1>^{d_2}$ and of the second one in the form
$[1,2]^{-d_3} [2,3]^{-d_1} [3,1]^{-d_2}$ one can see that if
\be
d_1 = h_1-h_2-h_3,~~~d_2 =h_2-h_3-h_1,~~~d_3=h_3-h_1-h_2,
\ee
then these polynomials represent true solutions.  For the three-point on-shell vertices therefore
we shall have \cite{Benincasa:2007xk}
\be\label{threlinearvertex}
M_3 = f <1,2>^{d_3} <2,3>^{d_1} <3,1>^{d_2} + k ~[1,2]^{-d_3} ~[2,3]^{-d_1} ~[3,1]^{-d_2},
\ee
where $f$ and $k$ are momentum independent constants.

The amplitude should vanish in the limit when all momenta $p^{\mu}$ become real,
in that case both spinor products $<i,j>$ and $[i,j]$ tend to zero. This means that if
$$
D=d_1 +d_2 +d_3 = -h_1-h_2 -h_3 >  0,
$$
then one should take $k=0$, while if
$$
-D=d_1 +d_2 +d_3 = -h_1-h_2 -h_3 <  0
$$
one should take $f=0$ in order to avoid singularities. Because
the dimensionality of spinors $\lambda$ and
$\tilde{\lambda}$ in formula (\ref{momentum}) is $[mass]^{1/2}$,  the dimensionality
of the three-point on-shell vertex  $M_3$  is $[mass]^D$.

The earlier investigation of the
three-point vertices in the light-front formulation of relativistic dynamics and
derivation of restrictions imposed on the helicities of scattered particles by the Poincar\'e group
were made in \cite{Bengtsson:1983pd,Bengtsson:1983pg,Bengtsson:1986kh}.  In covariant
formulation the interaction vertices were studied in \cite{Berends:1984rq,Berends:1985xx,Berends:1984wp}, see also
\cite{schwinger,fronsdal,Guttenberg:2008qe,fierz,pauli,rarita,singh,singh1,fronsdal1,
Curtright:1987zc,deser}.
The advantage of the spinor formulation is that it gives non-perturbative expressions.

In the generalized Yang-Mills theory \cite{Savvidy:2005fi,Savvidy:2005zm,Savvidy:2005ki,Savvidy:2010vb}
all interaction vertices
between high-spin fields have {\it dimensionless coupling constants in four-dimensional space-time}.
That is, the cubic interaction vertices have only first order derivatives and that the quartic vertices
have no derivatives at all.  Our
intension therefore  will be to find out constraints on helicities of the scattered particles, when the
amplitudes $M_3$ in (\ref{threlinearvertex})
have dimensionality of $mass$, that is,  $D=1$ and therefore the corresponding vertices  $\CV_3$
have a dimensionless coupling constants in four-dimensional space-time as it is the case in the generalized
Yang-Mills theory.

\section{\it Cubic Vertices with Dimensionless Coupling Constants }

The three-point on-shell vertex $M_3$ in (\ref{threlinearvertex}) will
have dimensionality of $mass$ if $D=1$.
In four dimensional space-time this will correspond to the three-particle interaction
vertex $\CV_3$, which has dimensionless coupling constant,
the property of the main importance
\cite{Savvidy:2005fi,Savvidy:2005zm,Savvidy:2005ki,Savvidy:2010vb}. This condition can be expressed
as
\be
D =d_1 +d_2 +d_3 = -h_1-h_2 -h_3 =1,
\ee
when $k=0$ and as
\be
D= -(d_1 +d_2 +d_3) =  h_1 + h_2 +h_3 =1,
\ee
when $f=0$.
We can now express the helicity of the third particle in terms of the
first two independent helicities. In the case $D =d_1 +d_2 +d_3 = -h_1-h_2 -h_3 =1$ we have
\beqa
&d_1 = h_1-h_2 -h_3 = h_1-h_2 +(h_1+h_2+1)= 2h_1 +1\nn\\
&d_2=h_2-h_1- h_3=h_2-h_1+(h_1+h_2+1)=2h_2+1\nn\\
&d_3= h_3-h_1 - h_2 = -(h_1+h_2+1)-h_1-h_2=-2h_1-2h_2-1\nn
\eeqa
and for the amplitude we shall get:
\be\label{dimensionone1}
M_3 = f <1,2>^{-2h_1 -2h_2 -1} <2,3>^{2h_1 +1} <3,1>^{2h_2 +1},~~~~h_3= -1 - h_1 -h_2.
\ee
For the second solution, when $D= -(d_1 +d_2 +d_3) =  h_1 + h_2 +h_3 =1$, we have
\beqa
&d_1 = h_1-h_2 -h_3 = h_1-h_2 -(-h_1-h_2+1)= 2h_1 -1\nn\\
&d_2=h_2-h_1- h_3=h_2-h_1-(-h_1-h_2+1)=2h_2 - 1\nn\\
&d_3= h_3-h_1 - h_2 = (-h_1-h_2+1)-h_1-h_2=-2h_1-2h_2+1\nn
\eeqa
and for the amplitude we shall get:
\be\label{dimensionone2}
M_3 = k [1,2]^{2h_1 +2h_2 -1} [2,3]^{-2h_1 +1} [3,1]^{-2h_2 +1},~~~~h_3= 1 - h_1 -h_2.
\ee
The formulas (\ref{dimensionone1}) and (\ref{dimensionone2})
give a general expression for the  amplitude $M_3$ in terms
of two independent helicities $h_1$ and $h_2$. It allows to choose any
 $h_1$ and $h_2$ and then find out $h_3$ for which the three-particle interaction
 vertex in four-dimensional space-time will have dimensionless coupling constants
 $f$ or $k$.
{\it We conclude that the general three-point on-shell vertices  with dimensionless coupling
constants are the vertices (\ref{dimensionone1}) and (\ref{dimensionone2})}.

The important conclusion which one can draw from the above result is that
if all three massless particles have the same spin $s_1 = s_2 = s_3=s$ then the only
vertex with dimensionless coupling constant will be the one with $s=1$.
This is the case of Yang-Mills theory where indeed the trilinear interaction
has dimensionless coupling constant. But already for spin 2 the
three-particle vertex has dimensional coupling constant as it is the case
in the general relativity.

Because (\ref{dimensionone1}) and (\ref{dimensionone2}) allow many other
solutions with nonequal spins $s_1 \neq s_2 \neq s_3$, we can conclude that
 there exist infinitely many trilinear interaction vertices between high spin
fields with dimensionless coupling constants as it is the case in the generalized Yang-Mills theory
 \cite{Savvidy:2005fi,Savvidy:2005zm,Savvidy:2005ki,Savvidy:2010vb}.
Indeed, let us first consider the Yang-Mills theory where all spins are equal to one
$s_1= s_2= s_3=1$. For $h_1=\pm 1,h_2=-1$ we get $h_3=\mp 1$ and the amplitude
(\ref{dimensionone1}) yields
\beqa
M^{a_1a_2a_3}_3(+1,-1,-1)= g_{111}~f^{a_1 a_2 a_3} {<2,3>^4 \over <1,2><2,3><3,1>},~~\nn\\
M^{a_1a_2a_3}_3(-1,-1,+1)= g_{111}~f^{a_1 a_2 a_3} {<1,2>^4 \over <1,2><2,3><3,1>},
\eeqa
where $f^{abc}$ are the structure constants of the internal gauge group G.
For $h_1=\pm 1,h_2=1$ we get $h_3=\mp 1$ and the amplitude (\ref{dimensionone2}) gives
\beqa
M^{a_1a_2a_3}_3(+1,+1,-1)= g_{111}~f^{a_1 a_2 a_3} {[1,2]^4 \over [1,2][2,3][3,1]}, \nn\\
M^{a_1a_2a_3}_3(-1,+1,+1)= g_{111}~f^{a_1 a_2 a_3} {[2,3]^4 \over [1,2][2,3][3,1]}.
\eeqa
We are interested in studying all possible dimensionless interactions between  spin 1 and spin 2
particles. There are no solutions with $s_1=1, s_2=1, s_3=2$, but in the generalized Yang-Mills theory
\cite{Savvidy:2005fi,Savvidy:2005zm,Savvidy:2005ki,Savvidy:2010vb,Savvidy:2008ks,Antoniadis:2009rd}
there are solutions
with one vector and two tensors $s_1=2, s_2=1, s_3=2$, they are:
\beqa
M^{a_1a_2a_3}_3(+2 ,-1 ,-2 )&=& g_{122}~f^{a_1 a_2 a_3} {<2,3>^6 \over <1,2>^3<2,3><3,1>},\nn\\
M^{a_1a_2a_3}_3(+2 ,+1,-2  )&=& g_{122}~f^{a_1 a_2 a_3} {[1,2]^6 \over [1,2][2,3]^3[3,1]}.
\eeqa
Thus there is a nontrivial
interaction between spin 1 and two spin 2 massless bosons with
dimensionless coupling constant exactly as in the generalized Yang-Mills theory
\cite{Savvidy:2005fi,Savvidy:2005zm,Savvidy:2005ki,Savvidy:2010vb}.
There exists a non-trivial cubic vertex of spin-1 and spin-s bosons
with second spin-s boson
\beqa\label{1ssvertex}
M^{a_1a_2a_3}_3(-s ,-1 ,+s )&=& g_{1ss}~f^{a_1 a_2 a_3} {<1,2>^{2s+2} \over <1,2> <2,3>^{2s-1}<3,1>},\nn\\
M^{a_1a_2a_3}_3(+s ,-1 ,-s )&=& g_{1ss}~f^{a_1 a_2 a_3} {<2,3>^{2s+2} \over <1,2>^{2s-1} <2,3> <3,1>},\nn\\
M^{a_1a_2a_3}_3(+s ,+1,-s  )&=& g_{1ss}~f^{a_1 a_2 a_3} {[1,2]^{2s+2}\over [1,2][2,3]^{2s-1}[3,1]},\nn\\
M^{a_1a_2a_3}_3(-s ,+1,+s  )&=& g_{1ss}~f^{a_1 a_2 a_3} {[2,3]^{2s+2}\over [1,2]^{2s-1}[2,3][3,1]},
\eeqa
as well as with spin-(s-2) and spin-(s+2) bosons
\beqa
M^{a_1a_2a_3}_3(-s ,+1 ,+s-2 )&=& g_{1ss-2}~f^{a_1 a_2 a_3} {<1,2>^{2s-3}<2,3>  <3,1>^3 \over   <2,3>^{2s} },\nn\\
M^{a_1a_2a_3}_3(+s ,-1 ,-s+2 )&=& g_{1ss-2}~f^{a_1 a_2 a_3} {[1,2]^{2s-3}[2,3][3,1]^3 \over  [2,3]^{2s}},\nn\\
M^{a_1a_2a_3}_3(+s ,+1,-s-2  )&=& g_{1ss+2}~f^{a_1 a_2 a_3} {<2,3>^{2s+2}\over <1,2>^{2s-3} <2,3><3,1>^3},\nn\\
M^{a_1a_2a_3}_3(-s ,-1,+s+2  )&=& g_{1ss+2}~f^{a_1 a_2 a_3} {[2,3]^{2s+2} \over [1,2]^{2s-3}[2,3][3,1]^3}.
\eeqa
With these vertices in hand we can compute the gluon fusion amplitudes into two high spin-s gauge bosons
$s=2,3,...$.
There are two on-mass-shell diagrams in Fig.\ref{fig1} which contribute to this process.

\section{\it Four-Particle Scattering Amplitudes }

\begin{figure}
\centerline{\hbox{\psfig{figure=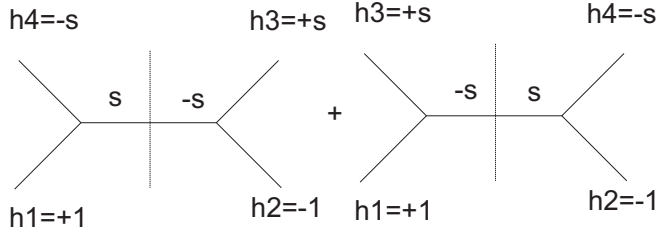,height=3cm,angle=0}}}
\caption[fig1]{The  BCFW diagrams for the process $V+V \rightarrow T+T$.}
\label{fig1}
\end{figure}

In this section, we intend to calculate the polarized cross sections for the reaction $V+V \rightarrow T+T$, to the lowest order
in cubic interaction coupling constants. There are two lowest-order diagrams contributing to
the annihilation process of a pair of vector bosons (gluons) into a pair of tensor gauge
bosons as shown in Fig.\ref{fig1}.
Vector gauge bosons  $V $ carry helicities $h_{1,2}=\pm 1$
and tensor gauge bosons $T$ carry helicities $h_{3,4}=\pm s$.

We shall choose to deform the momenta of the initial gluons  $p_1(z)=
(\lambda_1  + z \lambda_2) \tilde{\lambda}_1 $,~~ $p_2(z)=
\lambda_2(\tilde{\lambda}_2 - z\tilde{\lambda}_1) $ of helicities $h_1 =+1,~ h_2 =-1$
and shall leave the final tensor bosons momenta $p_3, p_4$  of helicities $h_3 =+s,~h_4 =-s$ undeformed.
The contribution of two diagrams takes the form
\beqa\label{firstseconddiagram}
M^{abcd}_4(+1,-1, +s,-s) =  2 i  ~g^2_{1ss}~f^{ade} f^{bce}~  {(-1)^{s+1} ~[1,3]^{s+1}~ <2,4>^{s+1}~
\over p^2_{12} ~p^2_{14}~ [1,4]^{s-1}~ <2,3>^{s-1}}+ \nn\\
+2i  ~g^2_{1ss}f^{ace} f^{bde}~{(-1)^{s+1} ~[1,3]^{s+1}~ <2,4>^{s+1}~
\over p^2_{12} ~p^2_{13}~ [1,4]^{s-1}~ <2,3>^{s-1}} ,
\eeqa
which can be written in a following factorized form:
\beqa\label{firstseconddiagram1}
M^{abcd}_4(+1,-1, +s,-s)
=2 i  ~g^2_{1ss} (-1)^{s+1} {  [1,3]^{s+1}  <2,4>^{s+1}
\over   [1,4]^{s-1} <2,3>^{s-1}}
 \{ { f^{ade} f^{bce} \over p^2_{12} ~p^2_{14}  }+
{   f^{ace} f^{bde} \over p^2_{12} ~p^2_{13} }  \}
\eeqa
where in the brackets we have a pure gluonic amplitude.
The alternative helicity amplitude  $M^{abcd}_4(+1,-1, -s,+s)$ can be found in a similar way.

It is important that a different choice of the momenta deformations will give the same
result. Let us  deform the momenta of the vector  and tensor bosons $p_1(z)=
(\lambda_1 +z \lambda_4) \tilde{\lambda}_1 $,~~ $p_4(z)=
\lambda_4(\tilde{\lambda}_4 - z\tilde{\lambda}_1) $ of helicities $h_1 =+1,~ h_4 =-s$
and leave the   momenta $p_2, p_3$  of helicities $h_2 =-1,~h_3 =+s$ undeformed.
The contribution of the first diagram  gives
\beqa\label{firstdiagram2defor}
 2 i  ~g_{111}g_{1ss}~f^{abe} f^{dce}~  { ~[1,3]^{2}~ <2,4>^{2s }~
\over p^2_{12}~  [3,2]~ <3,2>^{2s-1}},
\eeqa
multiplying it by $([1,3]  <1,4>/ [1,3] <1,4>)^{s-1}$ and combining different terms one can get
\beqa
 2 i  ~g_{111}g_{1ss}~f^{abe} f^{dce}~  { (-1)^{s+1}  ~~~([1,3]   <2,4>)^{ s+1 }~
\over p^2_{12}~p^2_{23}~  ([14]  <2,3>)^{s-1}}     \cdot  \left(   {[1,4]   <2,4>
\over  [3,1] <2,3> } \right)^{s-1}  .           \nn
\eeqa
From the momentum conservation it follows that the ratio in the last brackets is equal to one.
Using the identity $f^{abe} f^{dce} = + f^{ade} f^{bce}-f^{ace} f^{bde}$
one can see that the  term with $f^{ade} f^{bce}$ of this
amplitude coincides with the first term in (\ref{firstseconddiagram}) if the coupling
constant $g_{1ss}$  coincides with the  Yang-Mills coupling constant $g_{YM}$
\be\label{couplingconstants}
g_{1ss} =  g_{111} \equiv g_{YM},~~~~~~s=1,2,3,....
\ee
The remaining piece of the amplitude from the first diagram is
\beqa\label{remainingpiece}
- 2 i  ~g_{111}g_{1ss}~f^{ace} f^{bde}~  { (-1)^{s+1}  ~~~([1,3]   <2,4>)^{ s+1 }~
\over p^2_{12}~p^2_{23}~  ([14]  <2,3>)^{s-1}}.
\eeqa
The second diagram gives the following contribution:
\beqa\label{seconddiagram2defor}
-2 i  ~g_{1ss}^2~f^{ace} f^{bde}~~   { (-1)^{s+1}  ~~~([1,3]   <2,4>)^{ s+1 }~
\over p^2_{13}~p^2_{14}~  ([14]  <2,3>)^{s-1}}     \cdot  \left(   {[3,4]   <2,3>
\over  [1,4] <1,2> } \right)^{s-1}
\eeqa
and from momentum conservation it follows that the ratio in the last brackets is equal to one.
The sum of this amplitude with the remaining piece (\ref{remainingpiece}) from the first diagram
gives
\beqa
2 i  ~g_{1ss}^2~f^{ace} f^{bde}~~   { (-1)^{s+1}  ~~~([1,3]   <2,4>)^{ s+1 }~
\over   ([14]  <2,3>)^{s-1}}     (  -{1\over p^2_{13}~p^2_{14}}-{1\over p^2_{12}~p^2_{23}})        \nn
\eeqa
and  coincides with the second term in (\ref{firstseconddiagram}).

The third possibility is to  deform the momenta of the tensor  and
vector bosons  $p_3(z)=
(\lambda_3 +z \lambda_4) \tilde{\lambda}_3 $,~~ $p_2(z)=
\lambda_2(\tilde{\lambda}_2 - z\tilde{\lambda}_3) $ of helicities $h_3 =+s,~ h_2 =-1$
and  leave the   momenta $p_1, p_4$   of helicities $h_1 =+1,~h_2 =-s$ undeformed.
The contribution of the first diagram gives
\beqa\label{firstdiagram3defor}
-2 i  ~g_{111}~g_{1ss}~f^{cde} f^{bae}~ \frac{<2,4>^{2s-2 }}{<2,3>^{2s-2 }}\frac{<2,4>^{4 }}{<1,2><2,3><3,4><4,1>}.
\eeqa
This is exactly the contribution of the first diagram of the previous deformation  \eqref{firstdiagram2defor}, because
using momentum conservation one can derive the following identities
$$
{[1,3] \over [1,2]}=-{<4,2> \over <4,3>},~~~~{[1,3]\over [2,3]}=-{<2,4> \over <1,4>}
$$
and then plugging them into the equation \eqref{firstdiagram2defor} we shall get \eqref{firstdiagram3defor}.
Similarly, the second diagram yields
\beqa\label{seconddiagram3defor}
-2 i  ~g^2_{1ss}~f^{cae} f^{bde}~ \frac{<2,4>^{2s-2 }}{<2,3>^{2s-2 }}\frac{<2,4>^{4 }}{<3,1><4,1><2,3><2,4>}.
\eeqa
Using the identity $[1,3]/[1,4]=-<2,4>/<2,3>$ it is straightforward to see that this
result coincides with the contribution of the second diagram (\ref{seconddiagram2defor})
of the previous deformation.
We conclude that the third deformation gives the same result for the scattering amplitude
as previous two  deformations.

The important observation which follows from the above consideration
is that when we deform the momenta of gluons, the exchanged particles have
helicities $h=+s$ and $h=-s$, while when we deform the momenta of gluon and tensor bosons,
the exchanged particles have helicities
$h=1$ and $h=-s$, therefore, as we have seen, a full consistency between different kinematical channels
of the scattered particles exists only if the high spin coupling constants $g_{1ss}$ fulfill
the relations (\ref{couplingconstants}), that is, they all coincide with the
Yang-Mills coupling constant $g_{YM}$.

A second comment concerns the deformations performed above. In all three cases,
we have checked that the amplitude vanishes quickly enough as $z\rightarrow \infty$,
so that there is no contribution from the contour at infinity and that there is no
additional poles in the complex plane. The third comment concerns a
possible deformation of the tensor particles $p_3(z)=
(\lambda_3  + z \lambda_4) \tilde{\lambda}_3 $,~~ $p_4(z)=
\lambda_4(\tilde{\lambda}_4 - z\tilde{\lambda}_3) $  of helicities $h_3=+s, h_4=-s$
leaving the gluons momenta $p_1, p_2$  of helicities $h_3 =+1,~h_4 =-1$ undeformed.
Let us inspect the amplitude (\ref{firstseconddiagram1})
performing the above deformation $\lambda_3 \rightarrow
 \lambda_3  + z \lambda_4,~~ \tilde{\lambda}_4 \rightarrow
\tilde{\lambda}_4 - z\tilde{\lambda}_3 $. As one can see the amplitude develops
the higher order poles in the complex plane of the deformation parameter $z$ and the standard
recurrence prescription should be modified.

As we shall see below (see formulas (\ref{firstseconddiagramgeneralholomorphic}) and
(\ref{firstseconddiagramgeneralholomorphicf}))
the amplitudes given above can be recasted as purely holomorphic functions of the spinor
variables $\lambda_i$
and will be used in the last section to generalized the MHV
amplitudes (\ref{A5}) . The necessary condition for the n-particle amplitudes
to be holomorphic $\partial M^{(n)}  / \partial \tilde{\lambda}  =0$ can be found
using the equation (\ref{helicityequation}) and the fact that the amplitudes are
Lorentz  $J_{ab} M^{(n)}=0$ and scale $D \hat{M}^{(n)} =0$ invariant functions,
from which it follows that
${1\over 2}  \sum_i \lambda^a_i {\partial \over \partial \lambda^a_i} M^{(n)}
= -(n-4)M^{(n)}$\cite{Witten:2003nn}.
So that the amplitudes are holomorphic if particle helicities fulfill the equation
\be
h_1+h_2 +...+h_n =n-4
\ee
and are anti-holomorphic if
\be
h_1+h_2 +...+h_n =-n+4.
\ee

The exceptionally interesting non-MHV amplitude which involves two vector bosons and two tensor bosons
of the different spins  $s$ and $s-2$ is:
\beqa\label{seconddiagramsdifs}
 M^{abcd}_4(+1,-s, +1,s-2) =
 2 i  ~g^2 ( f^{ace} f^{bde}{<2,4>^{3}
\over  <1,2> <2,3> <3,1>} -~~~~~~~~~~~~~~~~~~~~~~~\\
- f^{ade} f^{bce} {<2,4>^{4}
\over  <1,2> <2,3>^2 <4,1> }       ) \times
 \left(  {<2,3> \over  <3,4>} \right)^{2s-1},\nn
\eeqa
where  $s=2,3,...$. This amplitude is holomorphic and is
of special interest because it has only {\it one particle of negative helicity}. In comparison, the n-gluon
tree amplitudes for $n \geq 4$ with all but one gluon of positive helicity vanish \cite{Parke:1986gb,Witten:2003nn}.
Thus the n-particle tree amplitudes in generalized Yang-Mills theory have more rich structure.
It is well known that the tree level n-gluon scattering
amplitudes with all positive helicities are also vanish \cite{Parke:1986gb,Witten:2003nn}.
In generalized Yang-Mill theory this statement remains true and can be
prove by induction. The 3-particle amplitudes in generalized Yang-Mills theory with all positive
helicities vanish (\ref{dimensionone1}) and (\ref{dimensionone2}). Suppose that
this is true for n particle amplitude, then the tree amplitude for n+1 particles is
a sum over terms constructed from products of two amplitudes of fewer particles multiplied
by a Feynman propagator. One of these fewer amplitudes is always with all positive helicity
particles and therefore vanish. {\it Thus in generalized Yang-Mills theory the tree level n-particle
scattering amplitudes with all positive helicities vanish $A^{(n)}(+,...,+)=0$, but tree amplitudes with one negative
helicity particle are already nonzero} (\ref{seconddiagramsdifs}).

Finally one can compute scattering amplitudes of four arbitrary non-Abelian tensor
gauge bosons of helicities $h_1,h_2,h_3, h_4$.
In the first diagram of Fig.\ref{fig1} the intermediate helicity is $h=1+h_2 +h_3= 1-h_1-h_4$
and in the second $h= 1+h_2 +h_4= 1-h_1-h_3$, thus
one should have  always
\be\label{generalcondition}
h_1+h_2+h_3+ h_4=0.
\ee
We can always choose $h_1 >0,~h_2 <0$, and
the total amplitude takes the form:
\beqa\label{firstseconddiagramgeneral}
 M^{abcd}_4(h_1,h_2, h_3,h_4) = ~~~~~~~~~~~~~~~~~~~~~~~~~\nn\\
 \nn\\
  =2 i  ~g^2 ~f^{ade} f^{bce}~
{(-1)^{h_2+h_3} ~[1,3]^{2h_1+h_2+h_3} <2,4>^{h_3-h_2} <3,4>^{2h_2+2}
\over p^2_{12} p^2_{14}~ [1,4]^{h_2+h_3} <2,3>^{h_2+h_3} [3,4]^{2h_1-2}}  \nn\\
\\
 +2i  ~g^2 f^{ace} f^{bde}~{(-1)^{-h_1-h_3} ~[1,3]^{h_1+h_3} <2,4>^{h_1+h_3} <4,3>^{2h_2+2}
\over p^2_{12} p^2_{13}~ [1,4]^{ h_3-h_1} <2,3>^{h_1 +2h_2+h_3} [4,3]^{2h_1-2}}.\nn
\eeqa
The conservation of momenta gives
\beqa\label{identities}
{[1,3] \over [1,4]}= -{<2,4> \over <2,3>},~~{[1,3] \over [3,4]}=- {<2,4> \over <1,2>},~~
{[3,4] \over [1,4]}= {<1,2> \over <2,3>},
\eeqa
and can be used to obtain {\it purely holomorphic expression}:
\beqa\label{firstseconddiagramgeneralholomorphic}
 M^{abcd}_4(h_1,h_2, h_3,h_4) =~~~~~~~~~~~~~~~~~~~~~~~~\nn\\
\nn\\
= -2 i  g^2 f^{ade} f^{bce}\left({ <24> \over <21>} \right)^{2 h_1} \left({<34> \over <32>} \right)^{2 h_2}
  \left( {<24> \over <23> }\right)^{2 h_3}   { <12><34> \over  <14><32> }  \nn\\
  \\
 -2i  ~g^2 f^{ace} f^{bde}
\left({ <24> \over <21>} \right)^{2 h_1} \left({<34> \over <32>} \right)^{2 h_2}
  \left( {<24> \over <23> }\right)^{2 h_3}   { <12><34> \over  <13><24> }   .\nn
\eeqa
Further it can be written in the following factorized form
\beqa\label{firstseconddiagramgeneralholomorphicf}
  M^{abcd}_4(h_1,h_2, h_3,h_4)
 = -2 i  g^2  { <24>^{2 h_1 +2 h_3 -4}~ <34>^{2 h_2 +2} \over <23>^{2 h_2 +2 h_3} ~ <12>^{2 h_1 -2} }
  \times  \\
  \nn\\
 \times \left( f^{ade} f^{bce}   { <24>^4  \over  \prod <ii+1>  }
 +f^{ace} f^{bde}   {<24>^4  \over  \prod_{3\leftrightarrow 4} <ii+1> } \right) ,\nn
\eeqa
where in the brackets is the MHV amplitude for the spin-1 gauge bosons times a factor
which is the contribution of the high spin gauge bosons.
We have to notice that the derivation of this amplitude in different channels is consistent
if the coupling constants fulfill the following relations:
\be
~g_{s_1,s_4,\vert 1-s_1+s_4\vert }=g_{s_1,s_4,\vert 1+s_1-s_4 \vert }=
g_{s_2,s_3,\vert 1-s_2+s_3\vert }=g_{s_2,s_3,\vert 1+s_2 - s_3\vert }=g_{YM}~.
\ee
In the next section we shall use these amplitudes to calculate the production cross sections
of non-Abelian tensor gauge bosons.

\section{{\it Tensor Bosons Production Cross Sections}}

\begin{figure}
\centerline{\hbox{\psfig{figure=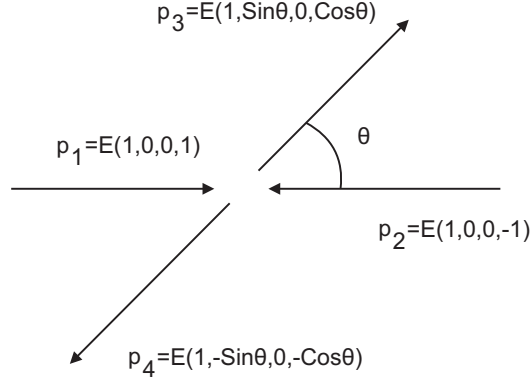,height=5cm,angle=0}}}
\caption[fig3]{The scattering $V+V  \rightarrow T+ T$, shown in the
center-of-mass frame. The $p_{1},p_2$ are momenta of the vector gauge bosons $V$ and $p_{3}, p_4$
are momenta of the tensor gauge bosons $T$.}
\label{fig3}
\end{figure}

The scattering process is illustrated in   Fig.\ref{fig3}.
Working in the center-of-mass frame, we make the following assignments:
$
p_{1}^{\mu}=E(1,0,0,1),~~ p_{2}^{\mu}=E(1,0,0,-1),
$~~and
$p_{3}^{\mu}=E(1,\sin\theta,0,\cos\theta),~~
p_{4}^{\mu}=E(1,-sin\theta,0,-\cos\theta),$
where $p_{1,2}$ are momenta of the vector bosons $V+V$ and $p_{3,4}$
- momenta of the tensor gauge bosons $T+T$. All particles are massless:
$p_{1}^{2} = p_{2}^{2} = p^{2}_3  = p^{2}_4  =0 $.
In the center-of-mass frame
the momenta satisfy the relations $\vec{p}_1  = -\vec{p}_2$, $\vec{p}_3  = -\vec{p}_4$.
The invariant variables of the process are:
\beqa
s =  2 (p_1 \cdot p_2),~~~
t=  -{s\over 2} (1-\cos \theta ),~~~
u= -{s\over 2} (1+\cos \theta ) \nn,
\eeqa
where $s= (2E)^2$ and $\theta$ is the scattering angle.
It is convenient to write the differential cross section in the center-of-mass frame with
tensor boson produced into the solid angle $d \Omega$ as
\be\label{crosssectionformula}
d\sigma = {1 \over 2 s} \vert M \vert^2 {1\over 32 \pi^2} d\Omega,
\ee
where the final-state density  is
$
d \Phi =   {1\over 32 \pi^2} d\Omega .
$
The spinor representation of momenta is
\beqa
&\lambda_1 = (\sqrt{2E},0),~~\lambda_2 = (0,\sqrt{2E}),~~\nn\\
&\lambda_3 = (\sqrt{E(1+\cos \theta )}, {E\sin\theta \over \sqrt{E(1+\cos \theta )}}),
\lambda_4 = ( {-E\sin\theta \over \sqrt{E(1+\cos \theta )}},\sqrt{E(1+\cos \theta )} ),\nn
\eeqa
and it allows to calculate all spinor invariant products in (\ref{firstseconddiagram}),(\ref{firstseconddiagram1}):
\beqa\label{polarizationRLRL}
M^{abcd}_4(+1,-1, +s,-s)
=i ~g^2 ~\left({ 1 -\cos\theta   \over  1 + \cos\theta    }\right)^{s-1} \times ~~~~~~~~~~~~~~~~~~~~~~~~~~~\nn\\
~\times[f^{ade} f^{bce}~ {(1 -\cos\theta )^{2}  \over (1 + \cos\theta )  }+
f^{ace} f^{bde}~ (1 -\cos\theta )],
\eeqa
as well as the alternative helicity amplitude
\beqa\label{polarizationRLLR}
M^{abcd}_4(+1,-1, -s,+s)
=i g^2 \left({ 1 +\cos\theta   \over  1 - \cos\theta    }\right)^{s-1}  \times~~~~~~~~~~~~~~~~~~~~~~~~~~~\nn\\
~\times[f^{ade} f^{bce}~ (1 +\cos\theta ) ~+
f^{ace} f^{bde}~{(1 +\cos\theta )^{2}  \over (1 - \cos\theta )  }  ].
\eeqa
To compute the cross section, we must square matrix elements
(\ref{polarizationRLRL}), (\ref{polarizationRLLR})
and then average over the symmetries of the initial bosons and sum over the
symmetries of the final tensor gauge bosons. This gives
\beqa\label{matrixelementA}
|\CM_{+-\rightarrow +s-s}|^{2}={g^{4}\over d(G)}C_{2}^{2}(G)
\biggl({1-\cos\theta \over  1+ \cos\theta}\biggr)^{2s-2}
\biggl({3+\cos^{2}\theta\over \sin^{4}\theta}\biggr)(1-\cos\theta)^{4}.
\eeqa
\beqa\label{matrixelementB}
|\CM_{+-\rightarrow -s+s}|^{2}={g^{4}\over d(G)}C_{2}^{2}(G)
\biggl({1+\cos\theta \over  1- \cos\theta}\biggr)^{2s-2}
\biggl({3+\cos^{2}\theta\over \sin^{4}\theta}\biggr)(1+\cos\theta)^{4},
\eeqa
where the invariant operator $C_2$ is defined by the equation
$ t^a t^a  = C_2~ 1 $ and $d(G)$ is the dimension of the internal group G.
Plugging squared matrix elements into the
cross-section formula  (\ref{crosssectionformula})
yields:
\beqa\label{crosssectionformulaLLRR}
d\sigma_{+-\rightarrow  +s-s}  =
  {\alpha^2 \over  s   }   ~{  C^2_2(G) \over   d(G) }~
 ~\biggl({1-\cos\theta \over  1+ \cos\theta}\biggr)^{2s-2}
\biggl({3+\cos^{2}\theta\over \sin^{4}\theta}\biggr)(1-\cos\theta)^{4}~ d\Omega ,
\eeqa
\beqa\label{crosssectionformulaRLLR}
d\sigma_{+-\rightarrow -s+s}  =
  {\alpha^2 \over  s   }   ~{  C^2_2(G) \over   d(G) }~
 ~\biggl({1+\cos\theta \over  1- \cos\theta}\biggr)^{2s-2}
\biggl({3+\cos^{2}\theta\over \sin^{4}\theta}\biggr)(1+\cos\theta)^{4}~ d\Omega ,
\eeqa
where
$
\alpha = {g^2_{YM} \over  4 \pi  }.
$
It is  instructive to compare the above cross sections
with the corresponding cross sections for the vector gauge bosons.
Indeed, the last formulas can be written in the factorized form stressing the
similarity with the gluon annihilation cross section
\beqa\label{crosssectionversus}
d\sigma_{ +s}  =
 ~\biggl({1-\cos\theta \over  1+ \cos\theta}\biggr)^{2s-2}
d\sigma_{ +1 }, ~~~~
d\sigma_{ -s}  =
 ~\biggl({1+\cos\theta \over  1- \cos\theta}\biggr)^{2s-2}
d\sigma_{-1},~~s=1,2,3,...,
\eeqa
where $\sigma_{+ -  \rightarrow +s-s} \equiv \sigma_{+s}$ and
$\sigma_{+ -  \rightarrow -s+s} \equiv \sigma_{-s}$.
It also shows explicitly the spin dependence of the cross
sections which has an amazingly simple form.
These cross sections have standards infrared singularity in
the forward and backward directions due to the massless character of the spectrum
of  non-Abelian tensor gauge bosons.
One can  sum the
production rate of high spin tensor gauge bosons. This gives
\be
\sum^{N}_{s=1}\biggl({1-\cos\theta \over  1+ \cos\theta}\biggr)^{2s-2}=
{(1+\cos\theta)^2 \over 4 \cos\theta } \biggl(1 + \biggl({1-\cos\theta \over  1+ \cos\theta}\biggr)^{2N-2}\biggr)
\ee
and taking the limit $N \rightarrow \infty$ one can get
\be
d\sigma_{tot} = \sum^{\infty}_{s=1}\biggl({1-\cos\theta \over  1+ \cos\theta}\biggr)^{2s-2} d\sigma_{+1}=
{(1+\cos\theta)^2 \over 4 \cos\theta } d\sigma_{+1},
\ee
which is valid in the region of the scattering angle
\be
\biggl({1-\cos\theta \over  1+ \cos\theta}\biggr) < 1~,~~~ 0  \leq \theta < \pi/2.
\ee
The characteristic feature of the total cross section is that it increases in the
transverse direction and tends to infinity already in the transversal plane $\theta \sim  \pi/2$.

In the general case of scattering of tensor gauge bosons of helicities  $h_1,h_2,h_3,h_4$ ~
($h_1+h_2+h_3+h_4=0,~h_1>0,h_2<0$),
one can also compute the cross section using the amplitude (\ref{firstseconddiagramgeneralholomorphic}),
(\ref{firstseconddiagramgeneralholomorphicf}). The cross section takes the form:
\beqa\label{crosssectiongeneral}
d\sigma_{h_1,h_2,h_3,h_4}  =
 {(1-\cos\theta)^{2h_1+2h_2+2h_3-2} \over 2^{2h_1-2h_2 -4} (1+ \cos\theta)^{2h_3-2} }
d\sigma_{ +1 }
 \eeqa
and {\it spectacularly falls exponentially as the spins of scattering particles increase}.

\section{\it Production of Tensor Gauge Bosons  and $(n-4)$ Jets }

In this section, we shall focus ourselves on color-ordered scattering amplitudes involving
two tensor particles of  helicity $+s$ and $-s$ respectively, one negative helicity gluon
and any number of gluons with positive helicity. In the case where $s=1$ one has
the MHV amplitudes for the scattering of vector bosons (gluons).
The expression for this amplitude is given by the
famous Parke-Taylor formula \cite{Parke:1986gb}.
We shall see that it is possible to write
a generalization of the Parke-Taylor formula for the case of arbitrary spin $s$.

We begin by giving the result which reads:
\be\label{park}
M^{(n)}(1^+,...i^-,...,k^{+s},....j^{-s},...,n^+)=
g^{n-2} \frac{<ij>^4}{\prod_{l=1}^{n} <l l+1>} \Big( \frac{<ij>}{<ik>}\Big)^{2s-2},
\ee
where $n$ is the total number of particles, and the dots stand for the
positive helicity gluons. Finally, $i$ is the position of the negative helicity gluon,
while $k$ and $j$ are the positions of the particles with helicities $+s$ and $-s$ respectively.
A first comment is that this expression is holomorhic in the spinors of the particles, exactly as
the MHV gluon amplitude.
A second comment is that for $s=1$ the second fraction in \eqref{park}
is absent and \eqref{park} reduces to the well-known result for the MHV amplitude.

We now proceed to prove \eqref{park} by induction.
For simplicity we shall look at the case in which the negative helicity gluon
sits at position 2 while the particles with helicities $s$ and $-s$ sit
at positions $n-1$ and $n$, respectively. During the proof,
it will become apparent that placing the particles at these special positions
plays no role and is primarily a matter of convenience.

Let us start with the case of five particles.
We shall deform the momentum of positive helicity gluon
sitting at position 1 and the negative helicity gluon sitting at position 2, as in section 5.
With this deformation there are two diagrams contributing.
The first diagram has three external particles attached to the left vertex
and two  external particles to the right vertex while the diagram on the right vice versa.
The right vertex of the first diagram gives a non-zero result only if the exchanged particle
has helicity $h>h_2+h_3=-1+1=0$. Thus, the only possibility is $h=1$. But for
$h=1$ the sum of the helicities of the left vertex is 2 and such a vertex
does not exist (a vertex with 4 particles is non-zero only if the sum of the helicities is zero (\ref{generalcondition})).
Consequently, the whole contribution to the five leg amplitude comes solely from the second
diagram. As before, the left vertex is non-zero only if $h>-(h_1+h_4)=s-1$, which means that
$h=s$.

The BCFW recursion relation allow us to write the amplitude as
\beqa\label{5-legs}
M^{(5)}=M^{(3)}(\hat{1},5,-\hat{P}_{51})~\frac{1}{P^2_{51}}~M^{(4)}(\hat{P}_{51},4,3,\hat{2}).
\eeqa
The expressions for $M^{(3)}(\hat{1},5,-\hat{P}_{51})$
and $M^{(4)}(\hat{P}_{51},4,3,\hat{2})$
are given by (\ref{1ssvertex})
\beqa\label{A3}
M^{(3)}(\hat{1},5,-\hat{P}_{51})=g \frac{[\hat{P}_{51}\hat{1}]^{2s+1}}{[5\hat{P}_{51}][\hat{1}5]^{2s-1}}
\eeqa
and (\ref{firstseconddiagramgeneralholomorphic}), (\ref{firstseconddiagramgeneralholomorphicf})
\beqa\label{A5}
M^{(4)}(\hat{P}_{51},4,3,\hat{2})=-g^2   \frac{<\hat{2}\hat{P}_{51}>^{4}}{\prod_{i\in R}<i i+1>}
\Big(\frac{<\hat{2}\hat{P}_{51}>}{<\hat{2}4>}\Big)^{2s-2},
\eeqa
where the product $\prod_{i\in R}<i i+1>$ is over the spinors of the right vertex of the diagram.
Plugging these expressions in \eqref{5-legs} we obtain
\beqa\label{5-legsint}
M^{(5)}= - g^3 \frac{<25>^4}{\prod_{i=1}^{5} <i i+1>} \times
\frac{[\hat{P}_{51}\hat{1}]^{2s-2}}{[\hat{1}5]^{2s-2}} \times
\frac{<2\hat{P}_{51}>^{2s-2}}{<2 4>^{2s-2}}.
\eeqa
The first fraction in \eqref{5-legsint} is the MHV amplitude for gluons and can be obtained by
isolating from \eqref{5-legs} the terms for $s=1$.
What remains are the last two fractions of \eqref{5-legsint}.
The next step involves using
\beqa\label{ids}
Y=\frac{[\hat{P}_{51}\hat{1}]}{[\hat{1}5]}
\frac{<2\hat{P}_{51}>}{<2 4>}=\frac{<2|P_{15}|1]/\bar{w}<2|P_{15}|1]/w}{[15]<24>}=
\frac{<2|P_{5}|1]<2|P_{5}|1]}{[15]<24>}\frac{1}{w\bar{w}},\nn
\eeqa
where
\beqa
w\bar{w}=[\hat{P}_{51}1]<2\hat{P}_{51}>=<2|P_{15}|1].
\eeqa
Plugging the last formula in $Y$ we get
\beqa
Y=-\frac{<25>}{<24>}.\nn
\eeqa
Altogether the five-particle amplitude becomes\footnote{The 4- and 5-particle amplitudes
correspond to 2-jet and 3-jet production in hadronic
collisions at very high energies, and so are of phenomenological importance.}
\beqa
M^{(5)}(1^+,2^-,3^{+},4^{+s},5^{-s})=  g^3 \frac{<25>^4}{\prod_{i=1}^{5} <i i+1>}\frac{<25>^{2s-2}}{<24>^{2s-2}}.
\eeqa

We are, now, ready to argue that the only contribution to the n-particle amplitude comes from
one diagram. The one which has a 3-particle vertex on the left part of the diagram and a ($n-1$)-particle
vertex on the right part. All other diagrams with different distribution of particles between the
left and right vertex give, in fact, zero. There are several ways to see this.

One way to get convinced is by considering the six-particle amplitude.
There are three diagrams which can potentially contribute.
The first one has a 3-particle vertex on the right. As above, this is non-zero only if $h>0$
which implies that $h=1$. But then the left vertex has particles with helicities $(1,-s,s,1,1)$ and
such a vertex does not exist.
The second diagram has two vertices with 4 particles each. The corresponding helicities are
$(1,-s,s,h)$ for the left vertex and $(-1,-h,1,1)$ for the right one. Thus $h$ can be $+1$ or $-1$. But
for $h=1$ the left vertex does not exist since the sum of the helicities is non-zero (\ref{generalcondition}).
Similarly, if the exchanged particle has helicity $h=-1$ it is the right vertex that does not exist.
Thus, the only contributing diagram is the last one.
One can go to seven or more particles and by using similar arguments
verify that there is only one contributing diagram.

Thus, there is only one diagram contributing to the $(1,-1,1,...,s,-s)$ amplitude.
It is the diagram where the left vertex is $(1,-s,s)$ while the right one is
$(-1,-s,s,1,...)$.
It is now obvious that the proof of the validity of \eqref{park} can be completed by induction.
The proof goes precisely as in the case for five particles. The only difference is that
one allows for $n-1$ particles in the right vertex. Then the only thing needed is to substitute
5 with $n$ and 4 with $n-1$ in \eqref{5-legs}, \eqref{5-legsint} and \eqref{ids}.

As we already notice the amplitude $M^{abcd}_4(+1,-s, +1,s-2)$ (\ref{seconddiagramsdifs}) is a first
example of non-MHV amplitude in generalized Yang-Mills theory.
With the amplitudes \eqref{park} in hands one can try to calculate
the n-particle amplitudes with more negative helicity particles
generalizing the calculation of \cite{Cachazo:2004kj,Georgiou:2004by,Georgiou:2004wu,Cachazo:2005ca}.

In conclusion we would like to remark that
one can perform a Fourier transformation of the holomorphic scattering amplitudes
\eqref{park} from momentum space to twistor space, in the spirit of \cite{Witten:2003nn}.
Since these amplitudes are holomorphic in the spinor variables this means that
they will be supported on an algebraic curve in twistor space of degree one and genus zero,
or in given case on a straight line.
Therefore one can  try to find out if there exists a tensionless string theory
\cite{Savvidy:2003fx,Savvidy:2005fe}, which will be the analog of the topological
B-model for the case of $N=4 SYM$, that can reproduce the tree level amplitudes of the
generalized Yang-Mills theory. It would be interesting to see if this possibility is realized.

We would like to thank F.~Cachazo for communication and remarks. One of us (G.S.) would like to thank
L.~J.~Dixon, T.~R.~Taylor and J.~M.~Drummond for helpful discussions.

\end{document}